\documentclass{emulateapj}

\def\kms{\ifmmode{\rm km\thinspace s^{-1}}\else km\thinspace s$^{-1}$\fi}
\def\vstar{V501\,Mon}

\shortauthors{Torres et al.}
\shorttitle{\vstar}

\begin{document}

\submitted{Accepted for publication in The Astronomical Journal}

\title{Absolute dimensions of the metallic-line eclipsing binary V501 MONOCEROTIS}

\author{
Guillermo Torres\altaffilmark{1},
Claud H.\ Sandberg Lacy\altaffilmark{2},
Kre\v{s}imir Pavlovski\altaffilmark{3},
Francis C.\ Fekel\altaffilmark{4}, \\ and 
Matthew W. Muterspaugh\altaffilmark{4,5}
}

\altaffiltext{1}{Harvard-Smithsonian Center for Astrophysics, 60
  Garden St., Cambridge, MA 02138, USA; e-mail:
  gtorres@cfa.harvard.edu}

\altaffiltext{2}{Department of Physics, University of Arkansas,
  Fayetteville, AR 72701, USA}

\altaffiltext{3}{Department of Physics, Faculty of Science, University
  of Zagreb, Bijeni\u{c}ka cesta 32, 10000 Zagreb, Croatia}

\altaffiltext{4}{Center of Excellence in Information Systems,
  Tennessee State University, Nashville, TN 37209, USA}

\altaffiltext{5}{Department of Mathematical Sciences, College of
  Science and Mathematics, Tennessee State University, Boswell Science
  Hall, Nashville, TN 37209, USA}

\begin{abstract}

We report extensive high-resolution spectroscopic observations and
$V$-band differential photometry of the slightly eccentric 7.02-day
detached eclipsing binary \vstar\ (A6m+F0), which we use to determine
its absolute dimensions to high precision (0.3\% for the masses and
1.8\% for the radii, or better). The absolute masses, radii, and
temperatures are $M_{\rm A} = 1.6455 \pm 0.0043~M_{\sun}$, $R_{\rm A}
= 1.888 \pm 0.029~R_{\sun}$, and $T_{\rm eff}^{\rm A} = 7510 \pm
100$~K for the primary and $M_{\rm B} = 1.4588 \pm 0.0025~M_{\sun}$,
$R_{\rm B} = 1.592 \pm 0.028~R_{\sun}$, and $T_{\rm eff}^{\rm B} =
7000 \pm 90$~K for the secondary. Apsidal motion has been detected, to
which General Relativity contributes approximately 70\%. The primary
star is found to be a metallic-line A star. A detailed chemical
analysis of the disentangled spectra yields abundances for more than a
dozen elements in each star. Based on the secondary, the system
metallicity is near solar: ${\rm [Fe/H]} = +0.01 \pm 0.06$. Lithium is
detected in the secondary but not in the primary. A comparison with
current stellar evolution models shows a good match to the measured
properties at an age of about 1.1 Gyr.

\end{abstract}


\section{Introduction}
\label{sec:introduction}

The eclipsing binary \vstar\ (with alternate names
2MASS\,J06404172$-$0106400 and TYC\,4799-1943-1; $V = 12.32$, SpT
A6m+F0 as determined here) was discovered by \cite{Wachmann:66} in a
photographic survey, and was given the original designation HBV\,444.
He proposed an orbital period of 7.021115~days, near the value we find
below, and found a displaced secondary minimum indicating an eccentric
orbit.  A reexamination of the original plates was done by
\cite{Bossen:72}, essentially confirming the original results, but
with better precision.  Since then, numerous times of eclipse have
been published, and \vstar\ has been mentioned as a target worthy of
followup for apsidal motion studies, possibly featuring an important
contribution to the precession from General Relativity
\citep{Gimenez:85, Gimenez:95}.  \cite{Sebastian:12} classified
\vstar\ as of spectral type \ion{A9}{3}. To date there has been no
spectroscopic study of the system.

In this paper we present new differential photometry of \vstar\ as
well as extensive high-resolution spectroscopy, the combination of
which enables us to derive accurate absolute dimensions for the
components as well as their chemical abundances for more than a dozen
elements. As we describe below, the primary is found to be a
metallic-line A star. Most such objects are members of binary systems
\citep{Abt:61, Abt:85}. Among the dozen or so Am binaries with well
determined properties, only a handful have been subjected to detailed
abundance analyses \citep[see][]{Lyubimkov:96, Torres:12,
  Pavlovski:14}; thus \vstar\ adds significantly to this class of
objects, and may assist in understanding the Am phenomenon.

\section{Eclipse timings}
\label{sec:timings}

Measurements of the times of eclipse for \vstar\ collected from the
literature are listed in Table~\ref{tab:timings}, and cover
approximately seven decades. An ephemeris curve solution using the
method of \cite{Lacy:92} gave clear indications of apsidal motion, as
suspected by \cite{Gimenez:85}, and resulted in a best fit eccentricity
of $e = 0.1313 \pm 0.0019$, longitude of periastron $\omega =
231\fdg54 \pm 0\fdg71$, and reference epoch of primary minimum HJD
$2,\!453,\!401.65029 \pm 0.00060$, where the inclination angle has
been held fixed at the value determined from our light-curve fit below
($i = 88\fdg022$). The apsidal period, however, is still very
uncertain ($U = 16,\!000 \pm 11,\!000$ yr), and strong correlations
are present between the eccentricity and the rate of apsidal motion in
this ``unconstrained'' solution. Given that the spectroscopic
measurements we present below provide a stronger handle on $e$, a
preferable approach is to perform a simultaneous fit of the eclipse
timings and the radial velocities, which should constrain the
parameters better. We defer a discussion of this solution until
Section~\ref{sec:orbit}.

\begin{deluxetable}{llcccc}
\tablewidth{0pc}
\tablecaption{Times of eclipse for \vstar.\label{tab:timings}}
\tablehead{
\colhead{HJD} &
\colhead{$\epsilon$} &
\colhead{} &
\colhead{} &
\colhead{$(O-C)$} &
\colhead{} \\
\colhead{(2,400,000$+$)} &
\colhead{(days)} &
\colhead{Eclipse} &
\colhead{Type} &
\colhead{(days)} &
\colhead{Source}}
\startdata
  30077.2220   &  \nodata  & 1  &    pg  &   $+$0.01762  &   1 \\ 
  30431.3800   &  \nodata  & 2  &    pg  &   $-$0.01654  &   1 \\ 
  31077.3560   &  \nodata  & 2  &    pg  &   $+$0.00819  &   1 \\ 
  31144.4230   &  \nodata  & 1  &    pg  &   $-$0.00467  &   1 \\ 
  31144.4280   &  \nodata  & 1  &    pg  &   $+$0.00033  &   1 \\ 
  31786.4970   &  \nodata  & 2  &    pg  &   $+$0.00703  &   1 \\ 
  31846.5810   &  \nodata  & 1  &    pg  &   $+$0.03275  &   1 \\ 
  33703.3080   &  \nodata  & 2  &    pg  &   $+$0.02785  &   1 \\ 
  33710.3090   &  \nodata  & 2  &    pg  &   $+$0.00764  &   1 \\ 
  34040.2580   &  \nodata  & 2  &    pg  &   $-$0.04021  &   1 \\ 
  34068.3530   &  \nodata  & 2  &    pg  &   $-$0.03005  &   1 \\ 
  34086.3060   &  \nodata  & 1  &    pg  &   $-$0.00692  &   1 \\ 
  34086.3400   &  \nodata  & 1  &    pg  &   $+$0.02708  &   1 \\ 
  34416.3070   &  \nodata  & 1  &    pg  &   $-$0.00259  &   1 \\ 
  34444.4320   &  \nodata  & 1  &    pg  &   $+$0.03758  &   1 \\ 
  34451.3920   &  \nodata  & 1  &    pg  &   $-$0.02362  &   1 \\ 
  34767.3870   &  \nodata  & 1  &    pg  &   $+$0.01711  &   1 \\ 
  34767.4250   &  \nodata  & 1  &    pg  &   $+$0.05511  &   1 \\ 
  34770.4820   &  \nodata  & 2  &    pg  &   $-$0.02199  &   1 \\ 
  34781.3920   &  \nodata  & 1  &    pg  &   $-$0.02030  &   1 \\ 
  34781.4220   &  \nodata  & 1  &    pg  &   $+$0.00970  &   1 \\ 
  34795.4210   &  \nodata  & 1  &    pg  &   $-$0.03371  &   1 \\ 
  35107.5130   &  \nodata  & 2  &    pg  &   $-$0.00905  &   1 \\ 
  35160.5080   &  \nodata  & 1  &    pg  &   $-$0.04941  &   1 \\ 
  35160.5390   &  \nodata  & 1  &    pg  &   $-$0.01841  &   1 \\ 
  52320.38766  &  0.0025   & 1  &    ccd &   $+$0.00322  &   2 \\ 
  52983.5215   &  0.0002   & 2  &    ccd &   $+$0.00009  &   3 \\ 
  53376.7091   &  0.0005   & 2  &    ccd &   $-$0.00004  &   4 \\ 
  53390.7530   &  0.0006   & 2  &    ccd &   $+$0.00144  &   4 \\ 
  53401.6502   &  0.0005   & 1  &    ccd &   $+$0.00006  &   4 \\ 
  53404.7933   &  0.0005   & 2  &    ccd &   $-$0.00068  &   4 \\ 
  53671.6010   &  0.0005   & 2  &    ccd &   $+$0.00106  &   5 \\ 
  54422.8674   &  0.0005   & 2  &    ccd &   $-$0.00195  &   6 \\ 
  54450.9531   &  0.0007   & 2  &    ccd &   $-$0.00109  &   6 \\ 
  54454.8302   &  0.0004   & 1  &    ccd &   $-$0.00081  &   6 \\ 
  54475.8934   &  0.0008   & 1  &    ccd &   $-$0.00123  &   6 \\ 
  54787.9741   &  0.0006   & 2  &    ccd &   $+$0.00185  &   6 \\ 
  54791.8490   &  0.0010   & 1  &    ccd &   $+$0.00011  &   6 \\ 
  55135.8866   &  0.0019   & 1  &    ccd &   $-$0.00137  &   7 \\ 
  55170.9960   &  0.0013   & 1  &    ccd &   $+$0.00200  &   7 \\ 
  55493.9715   &  0.0009   & 1  &    ccd &   $+$0.00203  &   7 \\ 
  55500.9939   &  0.0011   & 1  &    ccd &   $+$0.00323  &   7 \\ 
  56339.65867  &  0.00026  & 2  &    ccd &   $-$0.00088  &   8 \\ 
  56339.65954  &  0.00038  & 2  &    ccd &   $-$0.00001  &   8 \\ 
  56339.66015  &  0.00022  & 2  &    ccd &   $+$0.00060  &   8 \\ 
  56676.6771   &  0.0009   & 2  &    ccd &   $-$0.00050  &   9    
\enddata
\tablecomments{Measurement errors ($\epsilon$) are listed as
  published. ``Eclipse'' is 1 for a primary eclipse, 2 for a secondary
  eclipse. ``Type'' is pg for photographic measurements and ccd for
  more modern determinations. Uncertainties for the photographic
  measurements are estimated to be $\epsilon = 0.025$ days, as
  explained in Section~\ref{sec:orbit}.  $O\!-\!C$ residuals are
  computed from the combined fit described there.  Sources for the
  times of eclipse are:
(1) {\tt http://var2.astro.cz/EN/brno/eclipsing\_binaries.php};
(2) \cite{Brat:07};
(3) Measured from a minimum observed by M.\ Wolf (priv.\ comm.);
(4) \cite{Lacy:06};
(5) \cite{Zejda:06};
(6) \cite{Lacy:09};
(7) \cite{Lacy:11};
(8) \cite{Diethelm:13};
(9) \cite{Lacy:14}.}
\end{deluxetable}

\section{Spectroscopic observations and radial velocities}
\label{sec:spectroscopy}

Spectroscopic observations of \vstar\ were carried out with three
different instruments. They began at the Harvard-Smithsonian Center
for Astrophysics (CfA) in 2005 November, using the now decommissioned
Digital Speedometer \citep[DS;][]{Latham:92} mounted on the 1.5\,m
Tillinghast reflector at the Fred L.\ Whipple Observatory on Mount
Hopkins (AZ). Seven spectra were recorded through 2009 March with an
intensified photon-counting Reticon detector, and cover a narrow span
of 45\,\AA\ centered at 5190\,\AA\ (\ion{Mg}{1}\,b triplet). The
resolving power of this instrument was $R \approx 35,\!000$, and the
signal-to-noise ratios of the spectra range from 13 to 22 per
resolution element of 8.5~\kms. Dusk and dawn exposures of the sky
were taken nightly to monitor the velocity zero point.

Thirty seven additional spectra were gathered from 2009 November to
2015 February with the Tillinghast Reflector Echelle Spectrograph
\citep[TRES;][]{Furesz:08} on the same telescope. This bench-mounted,
fiber-fed instrument provides a resolving power of $R \approx
44,\!000$ in 51 orders over the wavelength span 3900--9100\,\AA. The
signal-to-noise ratios of the 37 spectra range from 8 to 56 per
resolution element of 6.8~\kms.  IAU radial-velocity standard stars
were observed each night.

Between 2011 October and 2015 February we also obtained 57 usable
spectra of \vstar\ with the Tennessee State University 2\,m Automatic
Spectroscopic Telescope (AST) and a fiber-fed echelle spectrograph
\citep{Eaton:07} at Fairborn Observatory in southeast Arizona. The
detector for these observations was a Fairchild 486 CCD, with
15\,$\mu$m pixels in a $4096 \times 4096$ format.  The spectrograms
have 48 orders ranging from 3800--8260\,\AA.  Because of the faintness
of \vstar\ ($V = 12.32$), we used a fiber that produced a spectral
resolution of 0.4\,\AA, corresponding to a resolving power of
$15,\!000$ at 6000\,\AA. See \cite{Fekel:13} for additional
information about the AST facility. Our spectra have typical
signal-to-noise ratios per resolution element of 40 at 6000\,\AA.

Radial velocities (RVs) from the CfA spectra were measured using the
two-dimensional cross-correlation technique TODCOR \citep{Zucker:94},
with synthetic templates taken from a large library of calculated
spectra based on model atmospheres by R.\ L.\ Kurucz
\citep[see][]{Nordstrom:94, Latham:02}, restricted to the
\ion{Mg}{1}\,b region. Template parameters (effective temperature,
surface gravity, metallicity, and rotational broadening) were selected
by running grids of cross-correlations as described by
\cite{Torres:02}. We adopted initial surface gravities of $\log g =
4.0$, close to our final values in Section~\ref{sec:dimensions}, and
solar metallicity.  Further experiments showed that while solar
composition is favored for the secondary (star B), a better match to
the observed spectra is achieved by adopting a metallicity for the
primary (star A) of ${\rm [Fe/H]} = +0.5$, which we have often seen in
the past in objects with chemical peculiarities typical of Am
stars. This was a first indication that the primary component of
\vstar\ may be metallic-lined. The temperatures for the primary and
secondary that maximize the cross-correlation coefficient averaged
over all exposures are 7280~K and 6950~K, although these are sensitive
to the adopted metallicity and surface gravity and could be biased if
the primary composition is anomalous.  Formal uncertainties for these
temperatures are 200~K each. The rotational broadenings were
determined to be $v \sin i = 17 \pm 1$~\kms\ and $11 \pm 2$~\kms,
respectively.  The heliocentric velocities we obtained from the DS and
TRES spectra were placed on a common frame and are listed in
Table~\ref{tab:rvCfA}. Those from the DS include corrections for
potential systematic effects due to the narrow spectral window
\citep[see][]{Latham:96} determined from numerical simulations as
described by \cite{Torres:97}.  These corrections are smaller than
0.5~\kms\ for both stars, and have little effect on the results
presented later.

\begin{deluxetable*}{lrccrcccc}
\tablewidth{0pc}
\tablecaption{Heliocentric radial velocity measurements of \vstar\ from CfA.
 \label{tab:rvCfA}}
\tablehead{
\colhead{HJD} &
\colhead{$RV_{\rm A}$} &
\colhead{$\epsilon_{\rm A}$} &
\colhead{$(O-C)_{\rm A}$} &
\colhead{$RV_{\rm B}$} &
\colhead{$\epsilon_{\rm B}$} &
\colhead{$(O-C)_{\rm B}$} &
\colhead{Orbital} &
\colhead{} \\
\colhead{(2,400,000$+$)} &
\colhead{(\kms)} &
\colhead{(\kms)} &
\colhead{(\kms)} &
\colhead{(\kms)} &
\colhead{(\kms)} &
\colhead{(\kms)} &
\colhead{phase} &
\colhead{Instrument}}
\startdata
    53690.9255  &  $-$84.96 &   0.64 &  $-$0.42 &  84.73 &   2.32 &  $+$3.61  &  0.2002 &  DS \\
    54048.9456  &  $-$82.16 &   0.47 &  $+$0.50 &  76.84 &   1.72 &  $-$2.16  &  0.1915 &  DS \\
    54070.9780  &  $-$76.65 &   0.39 &  $+$0.10 &  72.80 &   1.40 &  $+$0.47  &  0.3295 &  DS \\
    54136.8728  &   62.18 &   0.46 &  $-$0.45 & $-$85.06 &   1.66 &  $-$0.16  &  0.7146 &  DS \\
    54424.9270  &   60.25 &   0.37 &  $-$0.07 & $-$82.04 &   1.32 &  $+$0.25  &  0.7409 &  DS 
\enddata
\tablecomments{This table is available in its entirety in
  machine-readable and Virtual Observatory (VO) forms in the online
  journal. A portion is shown here for guidance regarding its form and
  content.}
\end{deluxetable*}

We determined the light ratio separately for the seven DS and the 37
TRES spectra following \cite{Zucker:94}, and obtained values of
$\ell_{\rm B}/\ell_{\rm A} = 0.58 \pm 0.06$ and $0.53 \pm 0.02$,
respectively, at a mean wavelength of 5190\,\AA. We point out,
however, that these values are also sensitive to the metallicity and
temperatures adopted for the cross-correlation templates, and may be
biased if the composition of the primary is peculiar (which could in
turn affect the temperature derived for this star). For example, using
a primary temperature of 7500~K, closer to the value found below,
reduces the light ratio from TRES to $\ell_{\rm B}/\ell_{\rm A} = 0.49
\pm 0.02$.  Although a template mismatch may affect the derived
temperatures and light ratio, experience has shown that it has little
effect on $v \sin i$ or on the radial velocities.

\cite{Fekel:09} gave a general explanation of the velocity measurement
from our Fairborn echelle spectra. For \vstar\ we have used our
solar-type line list and measured velocities for the lines in that
list that are just in the orders that cover the wavelength region
4920--7100\,\AA.  Our velocities were determined by fitting the
individual lines with rotational broadening functions
\citep{Sandberg:11} that allowed both the depth and width of the line
fits to vary. Our unpublished measurements of several IAU solar-type
velocity standards show that these Fairborn Observatory velocities
have a zero-point offset of $-$0.6~\kms\ when compared to the results
of \cite{Scarfe:10}. Therefore, 0.6~\kms\ has been added to each
velocity. We list the final values in Table~\ref{tab:rvFairborn}.

\begin{deluxetable*}{lrccrccc}
\tablewidth{0pc}
\tablecaption{Heliocentric radial velocity measurements of \vstar\ from Fairborn Observatory.
 \label{tab:rvFairborn}}
\tablehead{
\colhead{HJD} &
\colhead{$RV_{\rm A}$} &
\colhead{$\epsilon_{\rm A}$} &
\colhead{$(O-C)_{\rm A}$} &
\colhead{$RV_{\rm B}$} &
\colhead{$\epsilon_{\rm B}$} &
\colhead{$(O-C)_{\rm B}$} &
\colhead{Orbital} \\
\colhead{(2,400,000$+$)} &
\colhead{(\kms)} &
\colhead{(\kms)} &
\colhead{(\kms)} &
\colhead{(\kms)} &
\colhead{(\kms)} &
\colhead{(\kms)} &
\colhead{phase}}
\startdata
    55864.835 &   46.10 &   0.69 &  $-$0.36 &  $-$65.50  &  1.26 &  $+$0.61  &  0.8207 \\
    55928.777 &   14.40 &   0.69 &  $+$0.05 &  $-$31.70  &  1.26 &  $-$1.81  &  0.9277 \\
    55945.678 &  $-$74.40 &   0.69 &  $-$0.03 &   71.40  &  1.26 &  $+$1.21  &  0.3348 \\
    55949.894 &   13.40 &   0.69 &  $+$1.76 &  $-$25.70  &  1.26 &  $+$1.14  &  0.9353 \\
    55958.876 &  $-$86.60 &   0.69 &  $+$0.03 &   85.70  &  1.26 &  $+$1.68  &  0.2145 
\enddata
\tablecomments{This table is available in its entirety in
  machine-readable and Virtual Observatory (VO) forms in the online
  journal. A portion is shown here for guidance regarding its form and
  content.}
\end{deluxetable*}

From the solar-type line list for the AST spectra, the average
projected rotational velocities of components A and B are 16.5 and
12.4~\kms. These $v \sin i$ values have an estimated uncertainty of
1~\kms.

\section{Spectroscopic orbital solution}
\label{sec:orbit}

Separate orbital solutions using the CfA and Fairborn velocities are
listed in Table~\ref{tab:orbit}, and are generally consistent with
each other.  In order to optimally combine the information contained
in the radial velocities and in the eclipse timings, we carried out a
joint solution using the two RV data sets and the times of eclipse
from Section~\ref{sec:timings}, solving for the spectroscopic orbital
elements and the apsidal motion simultaneously. The prescription for
the latter is based on the algorithm of \cite{Lacy:92}.  The long
baseline of the eclipse timings helps to constrain the period, and the
spectroscopic measurements constrain the orbital eccentricity, which
is otherwise strongly correlated with the apsidal motion if using only
the timings to determine $d\omega/dt$. Formal measurement
uncertainties were used to establish weights, and adjusted iteratively
during the solution so as to achieve reduced chi-squared values near
unity, separately for the primary and secondary components and for
each type of observation. For the photographic times of eclipse that
have no published uncertainties our procedure assigned an average
error of 0.025~days.  Photoelectric/CCD timings with published errors
in Table~\ref{tab:timings} were scaled by factors of 1.7 for the
primary and 2.2 for the secondary.  The inclination angle was held
fixed according to the value determined from our light curve fit in
Section~\ref{sec:photometry}. Tests show that its small error
contributes negligibly to the uncertainties in all other adjusted
quantities.

\begin{figure}
\epsscale{1.15}
\plotone{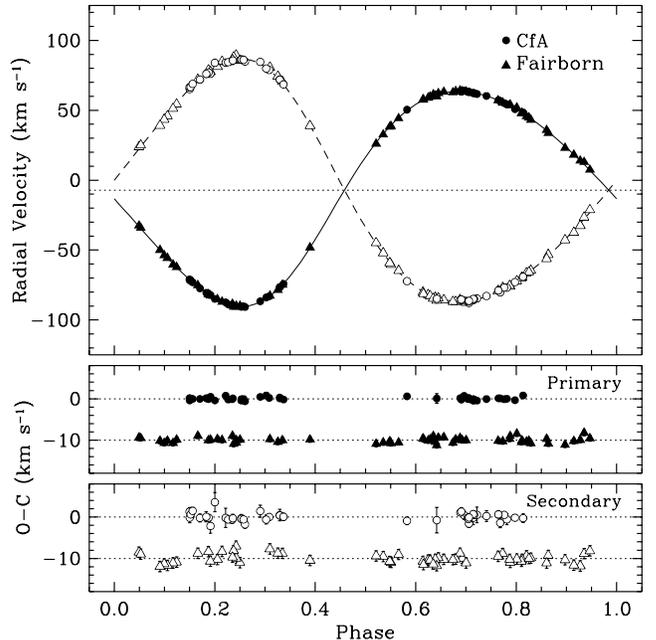}

\figcaption[]{Spectroscopic orbital solution for \vstar\ along with
  the measurements. Primary velocities are shown with filled symbols,
  and the dotted line represents the center-of-mass velocity. The
  bottom panels display the residuals, with those from Fairborn
  Observatory displaced vertically for clarity.\label{fig:orbit}}

\end{figure}

\begin{figure}
\epsscale{1.15}
\plotone{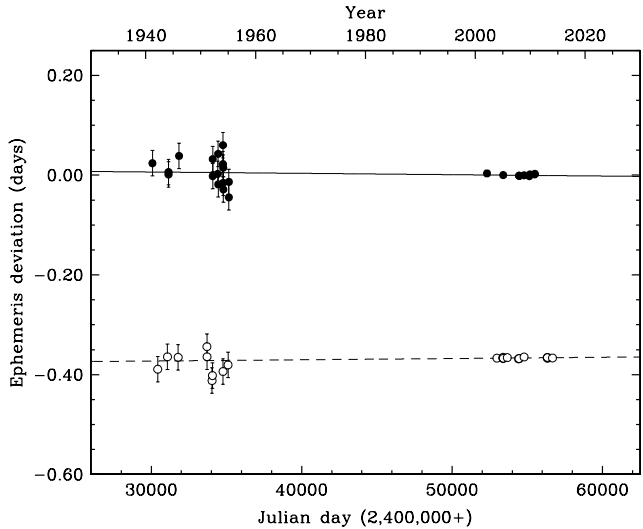}

\figcaption[]{Ephemeris curve for \vstar\ from our combined orbital
  fit, along with the eclipse timing measurements. Primary
  measurements are shown with filled circles.\label{fig:ephemeriscurve}}

\end{figure}

The results of our weighted least-squares fit are given in the last
column of Table~\ref{tab:orbit}.  Though improved over the preliminary
value listed in Section~\ref{sec:timings}, the apsidal motion period
$U$ remains relatively uncertain at $15400 \pm 8300$~yr. For each RV
set we included as an adjustable parameter an offset between the zero
points of the primary and secondary velocities, which for the CfA
measurements may arise because of a template mismatch, or for other
reasons. We also allowed for a possible overall shift between the
Fairborn and CfA velocities. All of these shifts turn out to be small,
but are statistically significant.  We illustrate the solution for the
spectroscopic elements in Figure~\ref{fig:orbit}, which shows the
measurements and their $O-C$ residuals from the fit, separately for
the Fairborn and CfA data sets. The residuals are listed in Tables~\ref{tab:rvCfA} and \ref{tab:rvFairborn}.
Those from the eclipse timings
and the best-fit ephemeris curve are displayed in
Figure~\ref{fig:ephemeriscurve}.

\begin{deluxetable*}{lccc}
\tablewidth{0pc}
\tablecaption{Spectroscopic/ephemeris-curve solutions for \vstar.\label{tab:orbit}}
\tablehead{
\colhead{\hfil~~~~~~~~~~~~~~~~~~~~Parameter~~~~~~~~~~~~~~~~~~~~} &
\colhead{CfA} &
\colhead{Fairborn} &
\colhead{RVs + timings}
}
\startdata
$P_{\rm sid}$ (days)\dotfill                                &           \nodata                  &           \nodata                  &    7.02120771~$\pm$~0.00000097 \\
$P_{\rm anom}$ (days)\dotfill                               &    7.021201~$\pm$~0.000014         &    7.021254~$\pm$~0.000061         &     7.0212164~$\pm$~0.0000053 \\
$\gamma$ (\kms)\dotfill                                     &     $-$6.91~$\pm$~0.11\phs         &    $-$6.692~$\pm$~0.099\phs        &      $-$7.067~$\pm$~0.045\phs \\
$K_{\rm A}$ (\kms)\dotfill                                  &      76.914~$\pm$~0.054\phn        &       76.91~$\pm$~0.12\phn         &        76.866~$\pm$~0.047\phn \\
$K_{\rm B}$ (\kms)\dotfill                                  &       86.58~$\pm$~0.13\phn         &       87.22~$\pm$~0.19\phn         &         86.71~$\pm$~0.11\phn \\
$e$\dotfill                                                 &      0.1359~$\pm$~0.0015           &      0.1348~$\pm$~0.0015           &       0.13388~$\pm$~0.00059 \\
$\omega_{\rm A}$ at $T_{\rm min~I}$ (deg)\dotfill           &      232.19~$\pm$~0.66\phn\phn     &      233.07~$\pm$~0.65\phn\phn     &        232.43~$\pm$~0.21\phn\phn \\
$T_{\rm min~I}$ (HJD$-$2,400,000)\dotfill                   &  53401.6532~$\pm$~0.0072\phm{2222} &   53401.633~$\pm$~0.025\phm{2222}  &   53401.65013~$\pm$~0.00044\phm{2222} \\
$T_{\rm peri}$ (HJD$-$2,400,000)\dotfill                    &   53404.224~$\pm$~0.012\phm{2222}  &   53404.227~$\pm$~0.057\phm{2222}  &    53404.2302~$\pm$~0.0039\phm{2222} \\
$d\omega/dt$ (deg~cycle$^{-1}$)\dotfill                     &          \nodata                   &           \nodata                  &       0.00045~$\pm$~0.00024 \\
$U$ (yr)\dotfill                                            &          \nodata                   &           \nodata                  &         15400~$\pm$~8300\phn \\
$i$ (deg)\dotfill                                           &          \nodata                   &           \nodata                  &           88.022 (fixed)     \\
$M_{\rm A} \sin^3 i$ ($M_{\sun}$)\dotfill                   &      1.6374~$\pm$~0.0049           &      1.6628~$\pm$~0.0078           &        1.6426~$\pm$~0.0043 \\
$M_{\rm B} \sin^3 i$ ($M_{\sun}$)\dotfill                   &      1.4545~$\pm$~0.0029           &      1.4663~$\pm$~0.0054           &        1.4562~$\pm$~0.0025 \\
$a_{\rm A} \sin i$ ($10^6$ km)\dotfill                      &      7.3570~$\pm$~0.0045           &       7.358~$\pm$~0.011            &        7.3545~$\pm$~0.0043 \\
$a_{\rm B} \sin i$ ($10^6$ km)\dotfill                      &       8.282~$\pm$~0.012            &       8.344~$\pm$~0.018            &         8.296~$\pm$~0.010 \\
$a \sin i$ ($R_{\sun}$)\dotfill                             &      22.481~$\pm$~0.019\phn        &      22.571~$\pm$~0.031\phn        &        22.498~$\pm$~0.016\phn \\
$q \equiv M_{\rm B}/M_{\rm A}$\dotfill                      &      0.8883~$\pm$~0.0013           &      0.8818~$\pm$~0.0023           &        0.8865~$\pm$~0.0012 \\
$\Delta{\rm RV}$ CfA (prim$-$sec) (\kms)\dotfill            &     $-$0.50~$\pm$~0.25\phs         &           \nodata                  &       $-$0.79~$\pm$~0.13\phs    \\
$\Delta{\rm RV}$ Fairborn (prim$-$sec) (\kms)\dotfill       &          \nodata                   &     $-$0.60~$\pm$~0.20\phs         &       $-$0.43~$\pm$~0.18\phs \\
$\Delta{\rm RV}$ (CfA$-$Fairborn) (\kms)\dotfill            &          \nodata                   &           \nodata                  &      $-$0.426~$\pm$~0.097\phs \\
$\sigma_{\rm A}$ (\kms)\dotfill                             &            0.26                    &             0.64                   &           0.28 / 0.64 \\
$\sigma_{\rm B}$ (\kms)\dotfill                             &            0.74                    &             1.08                   &           0.73 / 1.16 \\
Time span (days)\dotfill                                    &           3372.9                   &            1207.9                  &             26995.5 \\
$N_{\rm RV}$\dotfill                                        &             44                     &              57                    &             44 / 57   \\
$N_{\rm ecl}$\dotfill                                       &          \nodata                   &           \nodata                  &                46  
\enddata
\tablecomments{Symbols in the first column represent the sidereal and
  anomalistic periods, the center-of-mass velocity (which for the
  combined fit is on the CfA reference system), the velocity
  semi-amplitudes, eccentricity, longitude of periastron of the
  primary, reference time of primary eclipse, time of periastron
  passage, rate of apsidal motion, apsidal period, and inclination
  angle. The $\Delta{\rm RV}$ entries represent offsets between the
  primary and secondary velocities in each data set (primary minus
  secondary), and an overall difference in the velocity zero points in the
  sense CfA minus Fairborn. The physical constants used here are those
  adopted by \cite{Torres:10}.}
\end{deluxetable*}

\section{Differential photometry and light curve solution}
\label{sec:photometry}

An extensive program of CCD photometry was carried out using the NFO
WebScope \citep{Grauer:08} near Silver City, New Mexico, for the
purpose of gathering an accurate $V$-band light curve of \vstar\ for
analysis. A total of 6729 images were obtained over 281 nights between
2005 January and 2015 February. Comparison stars (TYC\,4799-1956-1 and
TYC\,4799-1098-1, with $V = 10.31$ and 11.21, respectively, both of
spectral type A0) within 10 arcmin in the same field of view as the
variable star were used to determine differential magnitudes. The
magnitudes were corrected for nightly variations in the photometric
zero point as we have done previously in similar studies \citep[see,
  e.g.,][]{Lacy:08}. The variable star differential magnitude was
referenced to the magnitude of the combined light of the two
comparison stars (variable minus comparisons) in each image.
Table~\ref{tab:NFO} lists these differential photometric measurements.

\begin{deluxetable}{@{~~~~~~~~}cc@{~~~~~~~~}}
\tablewidth{15pc}
\tablecaption{Differential $V$-band measurements of \vstar.
 \label{tab:NFO}}
\tablehead{
\colhead{~~~~~~HJD} &
\colhead{$\Delta V$~~~~~~~} \\
\colhead{~~~~~~(2,400,000$+$)} &
\colhead{(mag)~~~~~~~}}
\startdata
  53376.65788 &  2.672 \\
  53376.66208 &  2.704 \\
  53376.66422 &  2.697 \\
  53376.66633 &  2.708 \\
  53376.66839 &  2.726
\enddata
\tablecomments{This table is available in its entirety in
  machine-readable and Virtual Observatory (VO) forms in the online
  journal. A portion is shown here for guidance regarding its form and
  content.}
\end{deluxetable}

The $V$-band light curve of this well-detached system was analyzed
using the {\tt JKTEBOP} application of \cite{Southworth:11} \citep[see
  also][]{Popper:81, Etzel:81}. This program can combine the
brightness measurements with radial-velocity measurements and even
historical eclipse timings, but does not currently take into account
apsidal motion, nor can it use more than one radial-velocity set for
each component.  As mentioned earlier, the determination of
$d\omega/dt$ benefits greatly when the eccentricity can be constrained
by the spectroscopy. Consequently, we have elected to apply {\tt
  JKTEBOP} to only the brightness measurements of \vstar, and to treat
the radial velocities and eclipse timings separately, as already
described in Section~\ref{sec:timings} and Section~\ref{sec:orbit}.  The
fitted parameters from {\tt JKTEBOP} and their uncertainties are given
in Table~{\ref{tab:LC}, and are the central surface brightness of the
  secondary ($J_{\rm B}$) in units of the primary, the sum of the
  relative radii ($r_{\rm A}+r_{\rm B}$), the radius ratio ($k \equiv
  r_{\rm B}/r_{\rm A}$), the inclination angle ($i$), the linear
  limb-darkening coefficients ($u_{\rm A}$, $u_{\rm B}$), the
  eccentricity ($e$), the longitude of periastron of the primary
  ($\omega_{\rm A}$), the sidereal period ($P_{\rm sid}$), and a
  reference time of primary eclipse ($T_{\rm Min~I}$). The gravity
  darkening coefficients ($y_{\rm A}$, $y_{\rm B}$) were assumed to be
  the same for the two components and were held fixed at their
  theoretical values for radiative stars. The mass ratio was adopted
  from our spectroscopic orbit (Table~\ref{tab:orbit}). The normalized
  flux contributions of the stars ($\ell_{\rm A}$, $\ell_{\rm B}$)
  were computed internally by the program. Experiments in which we
  allowed for third light ($\ell_3$) returned values not significantly
  different from zero. The eccentricity and $\omega_{\rm A}$ from our
  fit are perfectly consistent with the much better determined values
  in Table~\ref{tab:orbit}. A graphical representation of the fitted
  model along with the observations appears in
  Figures~\ref{fig:LC}--\ref{fig:LCsec}.

\begin{deluxetable}{lc}
\tablewidth{0pc}
\tablecaption{Light-curve solution for \vstar.\label{tab:LC}}
\tablehead{
\colhead{~~~~~~~~~~~~Parameter~~~~~~~~~~~~} &
\colhead{Value}
}
\startdata
$J_{\rm B}$\dotfill                        &          0.738~$\pm$~0.040 \\
$r_{\rm A}+r_{\rm B}$\dotfill              &        0.15456~$\pm$~0.00057 \\
$r_{\rm A}$\dotfill                        &         0.0838~$\pm$~0.0013 \\
$r_{\rm B}$\dotfill                        &         0.0707~$\pm$~0.0013 \\
$k \equiv r_{\rm B}/r_{\rm A}$\dotfill     &          0.843~$\pm$~0.027 \\
$i$ (deg)\dotfill                          &         88.022~$\pm$~0.076\phn \\
$u_{\rm A}$\dotfill                        &           0.60~$\pm$~0.11 \\
$u_{\rm B}$\dotfill                        &           0.51~$\pm$~0.10 \\
$y_{\rm A} = y_{\rm B}$\dotfill            &               0.25 fixed \\
$e$\dotfill                                &         0.1296~$\pm$~0.0054 \\
$\omega_{\rm A}$ (deg)\dotfill             &          231.0~$\pm$~2.0\phn\phn \\
$\ell_3$\dotfill                           &                0 fixed \\
$\ell_{\rm A}$\dotfill                     &          0.648~$\pm$~0.020 \\
$\ell_{\rm B}$\dotfill                     &          0.352~$\pm$~0.023 \\
$\ell_{\rm B}/\ell_{\rm A}$\dotfill        &          0.544~$\pm$~0.036 \\
$P_{\rm sid}$ (days)\dotfill               &     7.02120688~$\pm$~0.00000050 \\
$T_{\rm Min~I}$ (HJD$-2,\!400,\!000$)\dotfill  &  53401.64906~$\pm$~0.00016\phm{2222} \\
$\sigma$ (mmag)\dotfill                    &              13.5980 \\
$N$\dotfill                                &                6729
\enddata
\end{deluxetable}

\begin{figure}
\epsscale{1.15}
\plotone{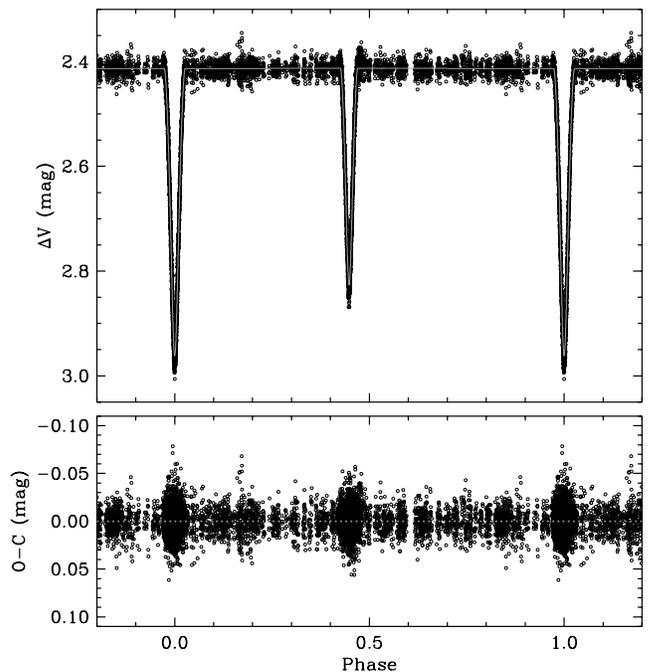}
\figcaption{NFO differential $V$-band observations of \vstar, shown
  with our best model fit. Residuals from the fit are shown at the
  bottom.\label{fig:LC}}
\end{figure}

\begin{figure}
\epsscale{1.15}
\plotone{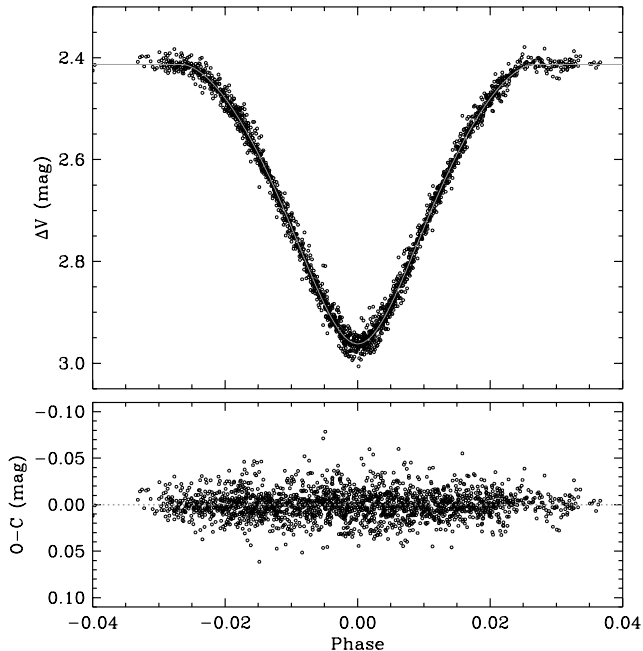}
\figcaption{Enlargement of Figure~\ref{fig:LC} around the primary
  minimum.\label{fig:LCprim}}
\end{figure}

\begin{figure}
\epsscale{1.15}
\plotone{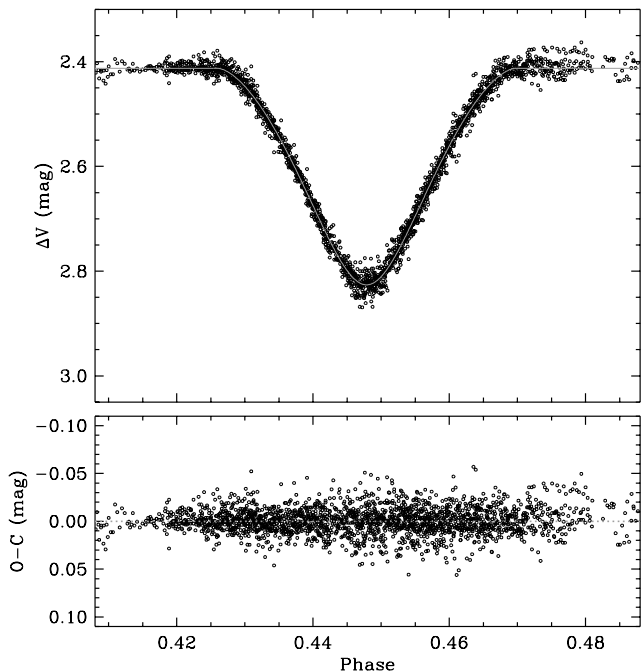}
\figcaption{Enlargement of Figure~\ref{fig:LC} around the secondary
  minimum.\label{fig:LCsec}}
\end{figure}

\section{Atmospheric parameters and chemical composition}
\label{sec:chemical}

In this section we describe our more detailed study of the atmospheric
properties of the \vstar\ components, including their elemental
abundances, that begins with a reconstruction of their individual
spectra by the method of spectral disentangling \citep{Simon:94}. The
method works best with a time series of high-resolution spectra more
or less evenly distributed in orbital phase \citep{Hensberge:08}, as
we have for \vstar.  In its most general form this procedure can solve
for the spectroscopic orbital elements in addition to the individual
spectra, so in a sense it is a generalization of the technique of
Doppler tomography \citep{Bagnuolo:91}, in which the separation of the
component spectra is made for a given set of known RVs for each star.
In this case the orbit is already well known from our fit in
Section~\ref{sec:orbit}, and the elements were held fixed, i.e.,
disentangling was performed in `pure separation' mode.

For this work we used the disentangling code {\tt FDBinary}
\citep{Ilijic:04}, which operates in Fourier space based on the
prescription of \cite{Hadrava:95}, with some improvements. The Fast
Fourier Transform implemented in {\tt FDBinary} gives more flexibility
in selecting the edges of the spectral segments for disentangling,
preserving the original spectral resolution. The choice of where to
place the edges of each segment is a critical step in Fourier-type
disentangling as these points should be strictly on the continuum.
Errors incurred in selecting the edges in spectra with a high density
of lines or with significant rotational broadening can add unwanted
undulations in the disentangled spectra of the components.

We ran {\tt FDBinary} on 27 of our TRES spectra of \vstar\ with the
highest signal-to-noise ratios (averaging about 22 per pixel), still
retaining good phase coverage.  Since imperfections in the
normalization of the echelle spectra are another potential source of
spurious undulations, we performed the disentangling in relatively
short spectral segments, usually 100--150\,\AA\ wide. The full
spectral range we analyzed is 4000--6760\,\AA, corresponding to about
30,500 pixels.

\begin{figure}
\epsscale{1.15}
\plotone{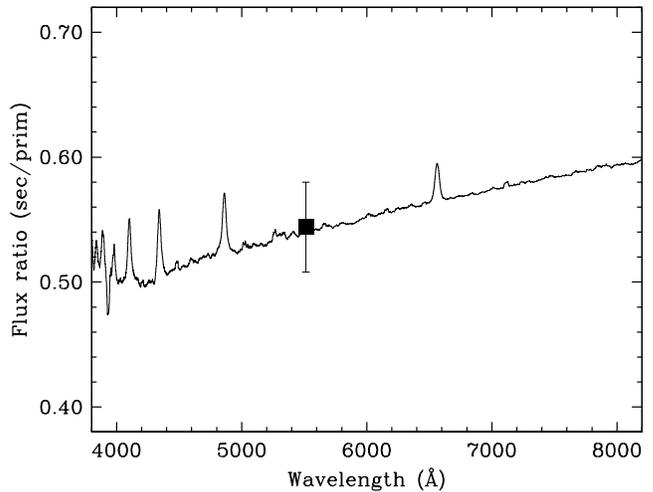}

\figcaption{Flux ratio for \vstar\ computed from synthetic spectra
  with parameters near those of the components ($T_{\rm eff} =
  7500$~K, $\log g = 4.0$, and ${\rm [Fe/H]} = +0.5$ for the primary,
  and $T_{\rm eff} = 7000$~K, $\log g = 4.0$, and ${\rm [Fe/H]} = 0.0$
  for the secondary), and the measured radius ratio $k = 0.843$. The
  $V$-band value estimated from our light curve fit ($\ell_{\rm
    B}/\ell_{\rm A}$; Table~\ref{tab:LC}) is also shown.
  \label{fig:fluxratio}}

\end{figure}

In the absence of a significant variation in the relative line
strengths of the components as a function of orbital phase there is an
ambiguity in the determination of the intrinsic line strengths for
each star. In other words, the disentangled spectra remain in the
common continuum of the total light of the binary system, and
renormalization to their individual continua requires external
information, such as light ratios inferred directly from the
light-curve fit of an eclipsing binary \citep[e.g.,][]{Hensberge:00,
  Pavlovski:05}. For \vstar\ we have this (Table~\ref{tab:LC}), but
effectively only at a single wavelength corresponding to the center of
the $V$ band.  In order to estimate the flux ratio at other
wavelengths we used synthetic spectra based on PHOENIX models
\citep{Husser:13} with parameters near those in Table~\ref{tab:absdim}
below, and we adopted the measured radius ratio $k$ from
Table~\ref{tab:LC}.  Figure~\ref{fig:fluxratio} shows the result,
which is in good agreement with the measured light ratio in $V$.
Therefore, for renormalization purposes we used a smoothed version of
this predicted relation.

Figure~\ref{fig:disentangling} shows a sample section of the
disentangled spectra for the primary and secondary of \vstar\ (second
and third from the top) on the common continuum of the composite
spectrum of the binary (top). Also shown at the bottom are the same
spectra renormalized according to the relation described above from
Figure~\ref{fig:fluxratio}. The signal-to-noise ratios of the
reconstructed spectra are 73 and 39 per pixel for the primary and
secondary, respectively, at a mean wavelength of 5500\,\AA.

\begin{figure}
\epsscale{1.15}
\plotone{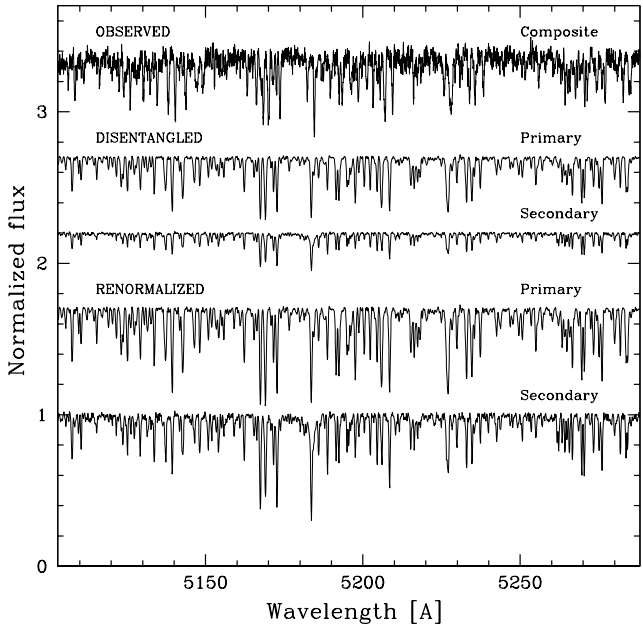}

\figcaption{Portions of the disentangled spectra of \vstar\ before and
  after normalization. From top to bottom the figure shows one of our
  composite spectra, the reconstructed primary and secondary spectra
  on the common continuum of the binary, and the same two spectra
  after normalization.\label{fig:disentangling}}

\end{figure}

Next we used these disentangled spectra to infer the stellar
parameters and abundances. Effective temperature and microturbulence
$\xi_{\rm t}$ were determined from the numerous iron lines, and the
usual conditions of excitation balance and null correlation of the
iron abundance with the reduced equivalent widths, respectively,
iterating as necessary. Equivalent widths of the \ion{Fe}{1} and
\ion{Fe}{2} lines carefully selected from the line list of
\cite{Bruntt:12} were measured with the {\tt UCLSYN} code
\citep{Smalley:01}, which was also used for the calculation of the
theoretical spectra. The surface gravities of the stars are well known
from our spectroscopic and light-curve solutions and were held fixed
at the values reported later in Table~\ref{tab:absdim}. Initial
results clearly indicated an enhanced metallicity for the primary
consistent with earlier indications of an Am nature, with ${\rm
  [Fe/H]} \sim +0.30$. We therefore adjusted our atmosphere model
accordingly, and repeated the analysis. The resulting excitation
temperatures and microturbulence velocities based on the \ion{Fe}{1}
lines are $T_{\rm eff} = 7540 \pm 140$~K and $\xi_{\rm t} = 3.1 \pm
0.1$~\kms\ for the primary, and $T_{\rm eff} = 6960 \pm 110$~K and
$\xi_{\rm t} = 1.3 \pm 0.1$~\kms\ for the secondary. The $\xi_{\rm t}$
value for the primary is consistent with results for normal A stars of
this temperature \citep[e.g.,][]{Kunzli:98, Gebran:10, Gebran:14},
while that of the secondary seems slightly lower than in other stars
of its type.

With both \ion{Fe}{1} and \ion{Fe}{2} lines being present in the
spectra of the \vstar\ components, the effective temperatures can in
principle also be derived from the condition of ionization balance, in
which the strength of the singly ionized Fe lines is influenced by
$\log g$ as well. However, with $\log g$ fixed as we described above,
the iron abundances from \ion{Fe}{1} and \ion{Fe}{2} in each star are
already in excellent agreement (and are A(\ion{Fe}{1}) $= 7.83 \pm
0.07$ and A(\ion{Fe}{2}) $= 7.80 \pm 0.08$ for the primary, and
A(\ion{Fe}{1}) $= 7.51 \pm 0.05$ and A(\ion{Fe}{2}) $= 7.50 \pm 0.07$
for the secondary)\footnote{We use the standard abundance notation in
  which ${\rm A(X)} = \log [n({\rm X})/n({\rm H})]+12$, where $n({\rm
    X})$ and $n({\rm H})$ are the numbers of atoms per unit volume of
  element X and of hydrogen.}, indicating no need to adjust the
$T_{\rm eff}$ values since they already satisfy ionization balance.

Additional estimates of the temperatures were derived by fitting the
wings of the H$\alpha$ and H$\beta$ line profiles with the {\sc
  starfit} code \citep{Tamajo:11, Kolbas:14}, which performs an
optimization based on a genetic algorithm \citep{Charbonneau:95}. For
this we used a precalculated grid of theoretical spectra in local
thermodynamical equilibrium \citep[for details, see][]{Kolbas:14,
  Kolbas:15}.  The metal lines superimposed on the wings were masked
out. The surface gravities and projected rotational velocities (see
below) were held fixed, and uncertainties in the derived temperatures
were calculated with a Markov Chain Monte Carlo technique (Pavlovski
et al., in prep.). The results from the H$\alpha$ lines are $7420 \pm
140$~K for the primary and $6790 \pm 180$~K for the secondary, and
from the H$\beta$ lines they are $7480 \pm 150$~K and $7070 \pm
160$~K, respectively. The latter results are consistent with our
earlier estimates from the iron lines, while the H$\alpha$ values are
somewhat lower, as expected from the shallower depth of formation of
this line.

Abundances for individual elements were derived using lines with the
most reliable $\log gf$ values, and are listed in
Table~\ref{tab:abundances}.  The quoted uncertainties take account of
the intrinsic scatter from the lines used in each case, as well as the
uncertainties in $T_{\rm eff}$, $\log g$, and $\xi_{\rm t}$. An
additional contribution has been added to the error budget
corresponding to a 1.5\% error in the light ratio of \vstar. The
results in Table~\ref{tab:abundances} are on the solar abundance scale
of \cite{Asplund:09}. The abundance pattern of the two components is
compared in Figure~\ref{fig:abundances}. While the secondary is seen
to have solar composition within the uncertainties, the pattern for
the primary is typical of Am stars, with a clear underabundance of Ca
and Sc, an enhancement of the iron group elements, and a large
overabundance of Ba and other heavy elements \citep{Preston:74}.

\begin{deluxetable*}{ll c ccc c ccc c c}
\tablewidth{0pc}
\tablecaption{Abundances from our disentangled TRES spectra of \vstar.\label{tab:abundances}}
\tablehead{
\colhead{} &
\colhead{} & &
\multicolumn{3}{c}{Primary} & &
\multicolumn{3}{c}{Secondary} & &
\colhead{} \\
\cline{4-6} \cline{8-10} \\ [-1ex]
\colhead{A} &
\colhead{X} & &
\colhead{Abundance} &
\colhead{[X/H]} &
\colhead{$N$} &&
\colhead{Abundance} &
\colhead{[X/H]} & 
\colhead{$N$} & &
\colhead{$\log\epsilon_{\sun}$}
}
\startdata
 11 & \ion{Na}{1} &&        \nodata    &       \nodata      &  \nodata &&  $6.28 \pm 0.05$ &  $+0.04 \pm 0.06$ &     6    &&  $6.24 \pm 0.04$ \\
 12 & \ion{Mg}{1} &&   $7.76 \pm 0.07$ &  $+0.16 \pm 0.08$  &     3    &&  $7.52 \pm 0.14$ &  $-0.08 \pm 0.15$ &     4    &&  $7.60 \pm 0.04$ \\
 14 & \ion{Si}{1} &&   $7.64 \pm 0.09$ &  $+0.13 \pm 0.09$  &    11    &&  $7.52 \pm 0.07$ &  $+0.01 \pm 0.08$ &     5    &&  $7.51 \pm 0.03$ \\
 20 & \ion{Ca}{1} &&        \nodata    &       \nodata      &  \nodata &&  $6.33 \pm 0.09$ &  $-0.01 \pm 0.10$ &     9    &&  $6.34 \pm 0.04$ \\
 20 & \ion{Ca}{2} &&   $5.85 \pm 0.08$ &  $-0.49 \pm 0.09$  &     6    &&      \nodata     &       \nodata     &  \nodata &&  $6.34 \pm 0.04$ \\
 21 & \ion{Sc}{2} &&   $2.52 \pm 0.08$ &  $-0.63 \pm 0.09$  &     3    &&  $3.11 \pm 0.04$ &  $-0.04 \pm 0.06$ &     6    &&  $3.15 \pm 0.04$ \\
 22 & \ion{Ti}{1} &&   $5.40 \pm 0.05$ &  $+0.45 \pm 0.07$  &    10    &&  $4.93 \pm 0.09$ &  $-0.02 \pm 0.10$ &    15    &&  $4.95 \pm 0.05$ \\
 22 & \ion{Ti}{2} &&   $5.31 \pm 0.06$ &  $+0.36 \pm 0.08$  &    11    &&      \nodata     &       \nodata     &  \nodata &&  $4.95 \pm 0.05$ \\
 24 & \ion{Cr}{1} &&   $6.04 \pm 0.09$ &  $+0.40 \pm 0.10$  &    15    &&  $5.63 \pm 0.06$ &  $-0.01 \pm 0.07$ &    44    &&  $5.64 \pm 0.04$ \\
 24 & \ion{Cr}{2} &&   $6.05 \pm 0.08$ &  $+0.41 \pm 0.09$  &     9    &&  $5.61 \pm 0.07$ &  $-0.03 \pm 0.08$ &    19    &&  $5.64 \pm 0.04$ \\
 25 & \ion{Mn}{1} &&   $5.60 \pm 0.11$ &  $+0.17 \pm 0.12$  &     7    &&  $5.36 \pm 0.06$ &  $-0.07 \pm 0.08$ &    10    &&  $5.43 \pm 0.05$ \\
 26 & \ion{Fe}{1} &&   $7.83 \pm 0.07$ &  $+0.33 \pm 0.08$  &   159    &&  $7.51 \pm 0.05$ &  $+0.01 \pm 0.06$ &   126    &&  $7.50 \pm 0.04$ \\
 26 & \ion{Fe}{2} &&   $7.80 \pm 0.08$ &  $+0.30 \pm 0.09$  &    18    &&  $7.50 \pm 0.07$ &  $+0.00 \pm 0.08$ &    14    &&  $7.50 \pm 0.04$ \\
 27 & \ion{Co}{1} &&        \nodata    &       \nodata      &  \nodata &&  $5.05 \pm 0.07$ &  $+0.06 \pm 0.10$ &    12    &&  $4.99 \pm 0.07$ \\
 28 & \ion{Ni}{1} &&   $6.69 \pm 0.08$ &  $+0.47 \pm 0.09$  &    41    &&  $6.22 \pm 0.07$ &  $+0.00 \pm 0.08$ &    36    &&  $6.22 \pm 0.04$ \\
 39 & \ion{Y}{2}  &&        \nodata    &       \nodata      &  \nodata &&  $2.17 \pm 0.08$ &  $-0.04 \pm 0.09$ &     6    &&  $2.21 \pm 0.05$ \\
 56 & \ion{Ba}{2} &&   $3.94 \pm 0.11$ &  $+1.76 \pm 0.14$  &     3    &&      \nodata     &       \nodata     &  \nodata &&  $2.18 \pm 0.09$
\enddata

\tablecomments{Columns list the atomic number, the element and
  ionization degree, the logarithm of the number abundance on the
  usual scale in which A(H) = 12, the logarithmic abundance relative
  to the Sun, and the number of spectral lines measured.  The last
  column gives the reference photospheric solar values from
  \cite{Asplund:09}.}

\end{deluxetable*}

The Li~6707\,\AA\ line is clearly visible in the spectrum of the
secondary star.  Line fitting yields an abundance estimate of ${\rm
  A(Li)} = 3.17 \pm 0.16$, in which the principal contribution to the
uncertainty is from the error in the effective temperature.  This
abundance is consistent with that expected for stars of this mass and
temperature \citep[see, e.g.,][]{Ramirez:12}. On the other hand, we
find no trace of the Li line in the primary star spectrum. We can only
place a rough upper limit of about ${\rm A(Li)} = 2.80$.

\begin{figure}
\vskip 10pt
\epsscale{1.15}
\plotone{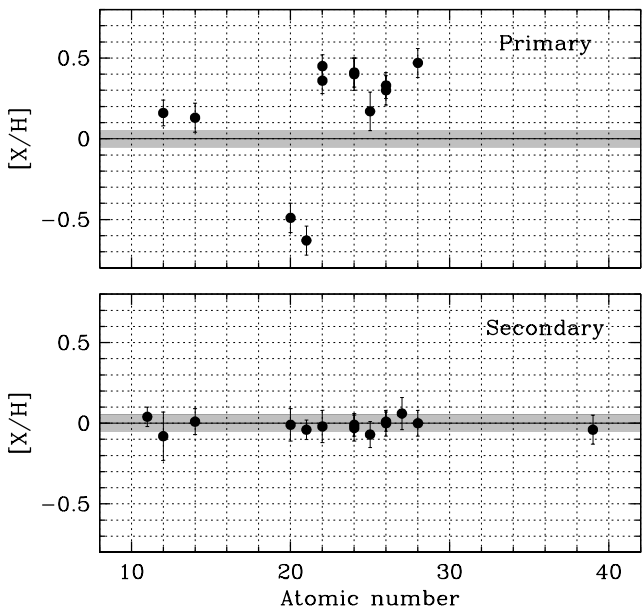}

\figcaption{Measured abundances for the primary and secondary of
  \vstar. The abundance of \ion{Ba}{2} for the primary star is not
  displayed, as it is far off the scale (+1.76, atomic number 56).
  For reference, the shaded area indicates the standard solar
  composition of \cite{Asplund:09}.\label{fig:abundances}}

\end{figure}

Finally, we also made an estimate of the macroturbulence velocities
($\zeta_{\rm RT}$) and projected rotational velocities from the
disentangled spectra. For the primary we obtained $\zeta_{\rm RT} =
6.3 \pm 1.6$~\kms\ and $v \sin i = 18.4 \pm 1.0$~\kms, and for the
secondary $\zeta_{\rm RT} = 5.3 \pm 1.2$~\kms\ and $v \sin i = 15.4
\pm 0.9$~\kms.

\section{Absolute dimensions}
\label{sec:dimensions}

The combination of the spectroscopic and light-curve elements yields
absolute masses for \vstar\ with relative errors of only 0.3\% and
0.2\% for the primary and secondary, and radius errors of 1.5\% and
1.8\%, respectively. Of the temperature determinations in the
preceding section based on the disentangled TRES spectra, we adopt for
each star a weighted average of the results from the metal lines and
from H$\beta$, which are $T_{\rm eff}^{\rm A} = 7510 \pm 100$~K and
$T_{\rm eff}^{\rm B} = 7000 \pm 90$~K. These correspond approximately
to spectral types of A6 and F0. The H$\alpha$ estimates are not
considered for the reason mentioned above. The measured
primary/secondary temperature difference, $\Delta T_{\rm eff} = 510
\pm 130$~K, is in very good agreement with the value $480 \pm 140$~K
inferred directly from the light curves through the fitted central
surface brightness parameter $J_{\rm B}$ and the visual absolute flux
calibration of \cite{Popper:80}.  We assume in the following that the
chemical composition of the system is represented by the measured
abundance for the secondary, given that the primary is anomalous.
Thus, ${\rm [Fe/H]} = +0.01 \pm 0.06$, based on the numerous
\ion{Fe}{1} lines.

The physical properties of the components are collected in
Table~\ref{tab:absdim}.  For the distance estimate we adopted an
apparent visual magnitude $V = 12.320 \pm 0.020$ from the AAVSO
Photometric All-sky Survey \citep[APASS;][]{Henden:12}, along with the
absolute luminosities and bolometric corrections from
\cite{Flower:96}.  The interstellar extinction toward \vstar, $A_V =
0.55 \pm 0.12$, was derived by iterations from the recent
three-dimensional reddening map of \cite{Green:15} based on Pan-STARRS
1 and 2MASS photometry, and assuming $A_V = 3.1 E(B-V)$.  The
resulting distance, $D = 1030 \pm 80$~pc, is very similar to an
independent calculation relying the visual surface fluxes
\citep[see][]{Popper:80}, which gives $D = 1050 \pm 70$~pc.

Table~\ref{tab:absdim} lists the predicted rotational velocities of
the stars under the assumption that tidal forces have synchronized
them with the orbital motion at periastron ($v_{\rm peri} \sin i$),
and the alternate assumption that they are pseudo-synchronized
\citep[$v_{\rm psync} \sin i$;][]{Hut:81}.\footnote{Both scenarios
  also assume the stars' spin axes are parallel to the orbital axis.}
Slight disagreements between our three estimates of $v \sin i$ from
Sections~\ref{sec:spectroscopy} and \ref{sec:chemical} prevent us from
distinguishing between the two possibilities for \vstar. In
particular, the two non-independent determinations from our CfA/TRES
spectra differ by about twice their combined errors for the secondary
component. Weight-averaging them, and then combining the result with
the Fairborn estimates for each star yields mean $v \sin i$ values of
$17.1 \pm 0.7$ and $12.8 \pm 1.2$~\kms, although we suspect systematic
errors are not negligible.

\begin{deluxetable}{lcc}
\tablewidth{0pt}
\tablecaption{Physical properties of \vstar.\label{tab:absdim}}
\tablehead{
\colhead{~~~~~~~~~Parameter~~~~~~~~~} &
\colhead{Primary} &
\colhead{Secondary}
}
\startdata
Mass ($M_{\sun}$)\dotfill                          & 1.6455~$\pm$~0.0043  &  1.4588~$\pm$~0.0025  \\
Radius ($R_{\sun}$)\dotfill                        &  1.888~$\pm$~0.029   &   1.592~$\pm$~0.028  \\
$\log g$ (cgs)\dotfill                             &  4.103~$\pm$~0.013   &   4.199~$\pm$~0.016  \\
Temperature (K)\dotfill                            &   7510~$\pm$~100\phn &    7000~$\pm$~90\phn\phn \\
$\log L/L_{\sun}$\dotfill                          &  1.007~$\pm$~0.027   &   0.743~$\pm$~0.043 \\
$BC_{\rm V}$\tablenotemark{a}\dotfill              &   0.03~$\pm$~0.10    &    0.03~$\pm$~0.10  \\
$M_{\rm bol}$ (mag)\tablenotemark{b}\dotfill       &  2.215~$\pm$~0.067   &    2.87~$\pm$~0.11   \\
$M_V$ (mag)\dotfill                                &   2.18~$\pm$~0.12    &    2.84~$\pm$~0.15  \\
$F_V$\tablenotemark{c}\dotfill                     & 3.8749~$\pm$~0.0057  &   3.846~$\pm$~0.010  \\
$A_V$ (mag)\dotfill                                & \multicolumn{2}{c}{0.55~$\pm$~0.12} \\
Distance (pc)\tablenotemark{d}\dotfill             & \multicolumn{2}{c}{1030~$\pm$~80\phn\phn} \\
$m-M$ (mag)\dotfill                                & \multicolumn{2}{c}{10.06~$\pm$~0.16\phn} \\
$v_{\rm peri} \sin i$\dotfill                      &   18.0~$\pm$~0.3\phn &    15.1~$\pm$~0.3\phn \\
$v_{\rm psync} \sin i$\dotfill                     &   15.1~$\pm$~0.2\phn &    12.7~$\pm$~0.2\phn \\
Measured $v \sin i$\,\tablenotemark{e}\dotfill     &   16.6~$\pm$~1.0\phn &    11.2~$\pm$~2.0\phn \\
Measured $v \sin i$\,\tablenotemark{f}\dotfill     &   18.4~$\pm$~1.0\phn &    15.4~$\pm$~0.9\phn \\
Measured $v \sin i$\,\tablenotemark{g}\dotfill     &   16.5~$\pm$~1.0\phn &    12.4~$\pm$~1.0\phn \\
$\xi_{\rm t}$ (\kms)\dotfill                       &    3.1~$\pm$~0.1     &    1.3~$\pm$~0.1 \\
$\zeta_{\rm RT}$ (\kms)\dotfill                    &    6.3~$\pm$~1.6     &    5.3~$\pm$~1.2 \\
${\rm [Fe/H]}$ (dex)\dotfill                       & \multicolumn{2}{c}{+0.01~$\pm$~0.06\phs} \\
${\rm A(Li~6707\,\AA)}$ (dex)\dotfill              &     $< 2.80$         &    3.17~$\pm$~0.16
\enddata
\tablenotetext{a}{Bolometric corrections from \cite{Flower:96}.}
\tablenotetext{b}{Uses $M_{\rm bol}^{\sun} = 4.732$ \citep[see][]{Torres:10}.}
\tablenotetext{c}{Visual absolute flux \citep{Popper:80}.}
\tablenotetext{d}{Relies on the luminosities and bolometric corrections.}
\tablenotetext{e}{Based on the original (composite) TRES spectra, determined from cross-correlation grids.}
\tablenotetext{f}{Based on the disentangled TRES spectra.}
\tablenotetext{g}{Based on the Fairborn spectra.}
\end{deluxetable}

\section{Comparison with stellar evolution models}
\label{sec:models}

Figure~\ref{fig:yale} presents a simultaneous comparison of the
masses, radii, and temperatures of the V501 Mon components with
stellar evolution models from the Yonsei-Yale series \citep{Yi:01,
  Demarque:04}. The solid lines correspond to evolutionary tracks for
the exact masses we measure, beginning at the zero-age main sequence
(ZAMS). The (solar-scaled) chemical composition of the models has been
adjusted to reproduce the observations, and the best match is achieved
for ${\rm [Fe/H]} = -0.07$, which is not far from the measured value
of ${\rm [Fe/H]} = +0.01 \pm 0.06$. An isochrone for this metallicity
and the best-fit age of 1.13~Gyr is shown with a dashed line. The
stars are seen to be slightly evolved from the ZAMS.  The radii and
temperatures are shown separately as a function of mass in
Figure~\ref{fig:mr}, and compared against the same best-fit model
isochrone. The agreement is very good.

\begin{figure}
\epsscale{1.15}
\plotone{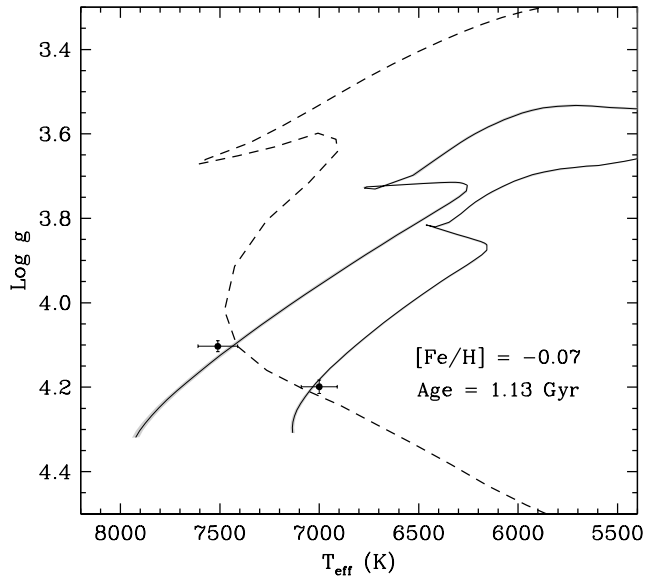}

\figcaption{Stellar evolution models from the Yonsei-Yale series
  compared against the observations for \vstar. Evolutionary tracks
  (solid lines) are shown for the exact masses we measure, and for the
  metallicity that produces the best match, which is near the measured
  value. The very small mass errors mean that the uncertainty in the
  placement of the tracks is very small (indicated in the plot with a
  barely visible shaded region around each track).  The best-fit
  isochrone shown by the dashed line has an age of
  1.13~Gyr.\label{fig:yale}}
\end{figure}

\begin{figure}
\epsscale{1.15}
\plotone{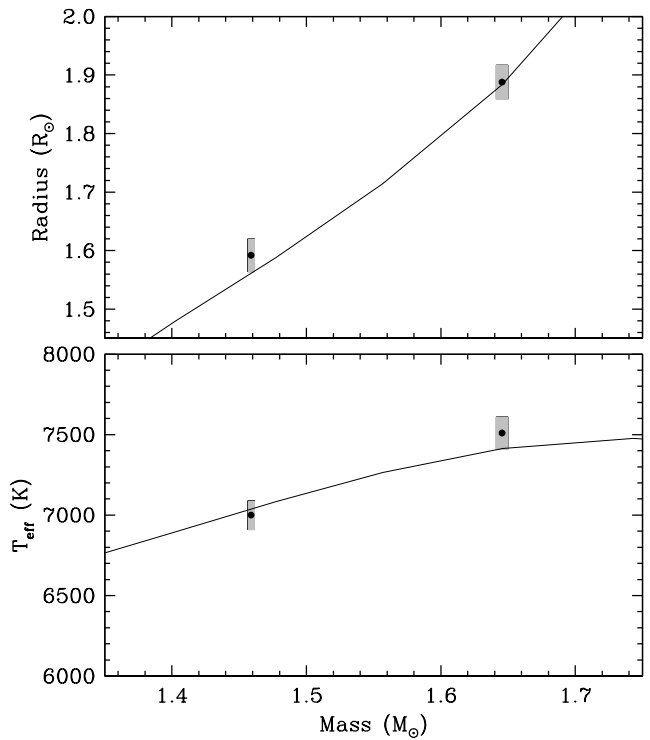}
\figcaption{Mass-radius and mass-temperature diagrams showing the
  reference isochrone from Figure~\ref{fig:yale} (${\rm [Fe/H]} = -0.07$,
  1.13~Gyr). The error boxes represent the observational uncertainties
  for \vstar.\label{fig:mr}}
\end{figure}

The difference between the best-fit metallicity from the models and
the measured composition of \vstar\ corresponds to about 1.3 times the
uncertainty.  If due to observational error in the temperatures and/or
composition, one possible source for the small discrepancy is a bias
in the light ratio we have adopted for the renormalization of the
disentangled TRES spectra.  The adopted dependence of the light ratio
with wavelength relies on synthetic spectra for normal stars from
\cite{Husser:13}, which for the primary may differ from the true
spectrum in subtle ways because of the Am nature of that star.

\section{Discussion and concluding remarks}
\label{sec:conclusions}

The high-precision masses and radii derived here for detached,
double-lined eclipsing binary \vstar\ place it in the elite group of
systems with the best determined properties
\citep[e.g.,][]{Torresetal:10}.  Our detailed chemical analysis adds
special value as fewer than three dozen of the systems with masses and
radii known to better than 3\% have been subjected to this type of
study.  The discovery that the primary is a metallic-line A star makes
it an important object that may help us understand more about this
phenomenon. So far as we are aware, only four other eclipsing binary
systems with similarly well measured masses and radii have been
published in which one or both components are Am/Fm stars, and which
also have detailed element-by-element abundance analyses: $\beta$\,Aur
\citep{Lyubimkov:96}, SW\,CMa and HW\,CMa \citep{Torres:12}, and
YZ\,Cas \citep{Pavlovski:14}.

Figure~\ref{fig:Am} shows all well-studied ($\sigma_M$ and $\sigma_R$
less than 3\%) A- and F-type stars in the field that are in the
temperature range of \vstar\ from the compilation of
\cite{Torresetal:10}, with additions from the recent literature.
Metallic-line stars are indicated with filled circles, and \vstar\ is
represented with squares. Numerous studies in the literature have
investigated the properties of Am/Fm stars in the field or in
clusters. A connection has been sought between abundance patterns in
these objects and various characteristics such as binary orbital
period or eccentricity, rotation, temperature, age, and others
\citep[e.g.,][and references therein]{Budaj:96, Budaj:97, Budaj:99,
  Kunzli:98, Abt:00, Fenovcik:04, North:04, Prieur:06}, but generally
without an accurate knowledge of the basic stellar properties (mass,
radius) or the age, which may be important if the anomalies depend
sensitively on them.  Only seven of the points in Figure~\ref{fig:Am}
(from \vstar\ and the four systems mentioned earlier) have detailed
photospheric abundances and many more may be needed before firm
correlations can begin to be discerned as a function of mass, radius,
or perhaps age.  For now, the figure shows that the currently
well-measured Am/Fm systems in the field are confined to the age range
from about 0.4~Gyr to 1.5~Gyr (indicated with dashed lines), whereas
normal A or F stars can be younger or older. The upper boundary also
corresponds to a roughly constant temperature of 6700--6800~K.
\cite{Kunzli:98} found similar limits for field stars. Younger Am
stars do of course exist in open clusters, such as the handful of
examples in the Pleiades \citep{Abt:78, Burkhart:97, Hui-Bon-Hoa:98,
  Gebran:08}, and perhaps even among pre-main-sequence stars such as
AK\,Sco \citep{Herbig:60, Andersen:89} with an age of only
$\sim$18~Myr \citep{Czekala:15}.

\begin{figure}
\vskip 15pt
\epsscale{1.15}
\plotone{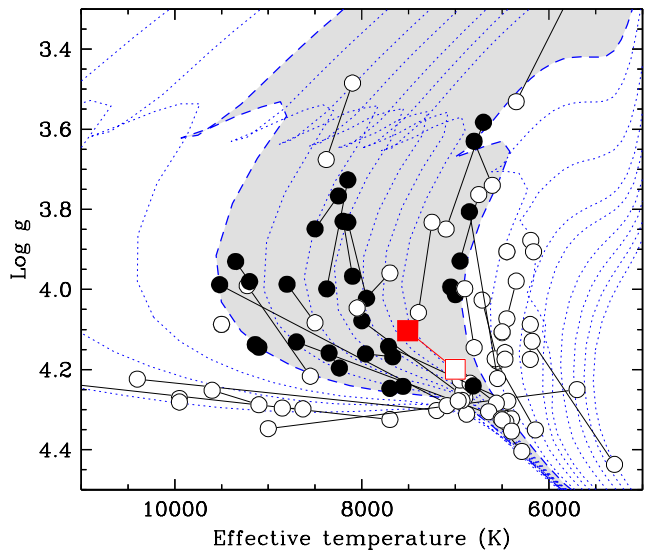}

\figcaption{Field eclipsing binaries (circles) with well determined
  masses and radii in the temperature range of
  \vstar\ (squares). Solid lines connect components of the same
  binary, and metallic-line stars are indicated with filled
  symbols. Solar-metallicity isochrones from the Yonsei-Yale series
  (0.1--13 Gyr) are shown with dotted lines for reference. The dashed
  lines delimit the age range 0.4--1.5~Gyr within which the Am/Fm
  stars in this sample fall.\label{fig:Am}}

\end{figure}

We detect Li in the spectrum of the secondary of \vstar\ but not in
the Am primary.  Some studies have suggested that the Li abundance in
Am stars is lower than in normal A stars by about a factor of 3
\citep[e.g.,][]{Burkhart:00}, while others have found no significant
difference \citep[e.g.,][]{Catanzaro:14}.  In the 1.1~Gyr system
\vstar\ the equal-age components do differ by at least a factor of 2,
though it is unclear whether this may be more related to the
temperature difference ($\Delta T_{\rm eff} \approx 500$~K in this
case).

Finally, the measurement of apsidal motion in eclipsing binary systems
has long been used to test models of the internal structure of stars
\citep[for a recent review, see][]{Claret:10}, specifically, the
degree of mass concentration towards the center. In addition to the
classical contributions to $d\omega/dt$ from tidal and rotational
distortions, in a few systems the contribution from General Relativity
\citep{Levi-Civita:37} is very important. This happens to be the case
for \vstar, for which we calculate $(d\omega/dt)_{\rm GR} =
0.00032$~deg~cycle$^{-1}$. Although the total measured apsidal motion
is still very uncertain ($d\omega/dt = 0.00045 \pm
0.00024$~deg~cycle$^{-1}$), the GR term appears to dominate, as
anticipated by \cite{Gimenez:85}, nominally contributing about 70\% to
the total amount. The internal structure constants for the primary and
secondary of \vstar\ from the models by \cite{Claret:04}, $\log k_2 =
-2.51$ and $-2.42$, lead to a predicted total apsidal motion rate
(including the GR term) of $(d\omega/dt)_{\rm tot} =
0.00046$~deg~cycle$^{-1}$ that is very close to the measured value.

\acknowledgments

We are grateful to P.\ Berlind, W.\ Brown, M.\ Calkins, G.\ Esquerdo,
D.\ Latham, A.\ Milone, R.\ Stefanik, S.\ Tang, and S.\ Quinn for help
in obtaining the CfA observations of \vstar\ with the DS and with
TRES, and to R.\ J.\ Davis and J.\ Mink for maintaining the CfA
echelle databases over the years. M.\ Wolf kindly provided his
photometric measurements of a secondary eclipse of \vstar. The authors
also wish to thank Bill Neely, who operates and maintains the NFO
WebScope for the Consortium, and who handles preliminary processing of
the images and their distribution.  GT acknowledges partial support
for this work from NSF grant AST-1509375. The work of KP has been
supported in part by the Croatian Science Foundation under grant
2014-09-8656. NSF grant 1039522 from the Major Research
Instrumentation Program, which was awarded to Tennessee State
University, enabled the acquisition of the Fairborn observations. In
addition, astronomy at Tennessee State is supported by the state of
Tennessee through its Centers of Excellence program.  This research
has made use of the SIMBAD and VizieR databases, operated at CDS,
Strasbourg, France, and of NASA's Astrophysics Data System Abstract
Service.


\end{document}